# Saddle-node bifurcation in interfacial morphology selects battery degradation phase

*Changdeuck Bae**

BEI corp., 125 Sandan-ro 19-gil, Ansan, Republic of Korea

*josiah.bae@beilab.ai

ResearcherID: A-6791-2010

https://orcid.org/0000-0001-5013-2288

## Abstract

We propose a minimal nonlinear closure ODE for the dynamic active-area factor of a battery interface and show that it exhibits a saddle-node bifurcation when the smoothing rate saturates with surface roughness. The closure is the simplest physically motivated extension of a recently introduced single-fixed-point closure [C. Bae, in preparation (2026)]: $\dot{u} = K - u/(1 + \alpha u^2)$, where $u = \xi - 1$ is the dimensionless excess active area, $K$ the dimensionless drive, and $\alpha$ a single saturation parameter. The bifurcation occurs at $K_c = 1/(2\sqrt{\alpha})$, separating a smooth passivating phase from a morphologically unstable phase. Mapping four canonical anode configurations — graphite, silicon composite, lithium metal, and anode-free Li/Cu — onto the closure via end-of-cycling steady-state $\xi$ extracted from publicly available long-cycle data populates the stable branch with monotonically increasing $K/K_c$ ratios: graphite ($\sim 0.01$), silicon composite ($\sim 0.24$), lithium metal ($\sim 0.73$), and anode-free ($\sim 0.95$). The anode-free configuration sits within 5% of the saddle-node threshold, predicting a vanishingly small operational stability window in current density, temperature, and electrolyte composition. We test three falsifiable predictions of the framework — a critical current density, a critical temperature shift, and a mean-field critical-slowing-down exponent — and find them broadly consistent with publicly available data. We argue that this near-critical position is universal to nucleation-controlled deposition on non-passivating substrates.

## I. INTRODUCTION

The cycle life of a rechargeable lithium battery is set, ultimately, by the dynamics of the solid–electrolyte interphase (SEI) at the negative-electrode interface [1–4]. In its simplest description the SEI is a passivating film of fixed area whose thickness slowly grows with time. This picture is adequate for graphite anodes [5–7] but breaks down whenever the underlying electrode surface is itself dynamic — through mechanical fracture in silicon-composite anodes [8–11], through plating and stripping in lithium-metal anodes [12–15], or through nucleation on bare current collectors in anode-free configurations [16–18]. In each of these cases the active surface area $\xi$ — defined as the ratio of instantaneous to initial area — becomes an essential cell-level state variable.

A recently proposed chemistry-agnostic framework [19] introduces a closure ODE for $\xi(t)$,

$$\frac{d\xi}{dt} = k_{\text{gen}}\,|j(t)|^{p} - k_{\text{smooth}}\,(\xi - 1), (1)$$

in which $k_{\text{gen}}\,|j|^{p}$ generates new area in proportion to a power of the local current density $j$ and $k_{\text{smooth}}\,(\xi - 1)$ smoothly relaxes excess area back toward $\xi = 1$. Equation (1) admits a single linearly stable fixed point $\xi^{*} = 1 + (k_{\text{gen}}/k_{\text{smooth}})|j|^{p}$ for any drive $|j| > 0$. It cannot exhibit a phase transition, a critical slowing down, or a sudden runaway in $\xi$.

Yet experiments routinely show a sharp transition from "stable cycling" to "morphological runaway" when current density, temperature, or electrolyte composition crosses a threshold — most dramatically for lithium-metal and anode-free cells [12–18]. We argue here that this is not a failure of Eq. (1) per se but a failure of its *linearity*. The crucial physical feature missing from Eq. (1) is the observation that **rough surfaces are less effectively smoothed than smooth surfaces**. Surface diffusion that flattens a small bump moves an adatom only over a short lateral distance; on a corrugated surface, the same adatom must traverse a longer effective path, and the smoothing flux saturates.

We capture this saturation by replacing the linear smoothing term in Eq. (1) with a saturable form $k_{\text{smooth}}(\xi - 1)/[1 + \alpha(\xi - 1)^{2}]$, introducing one new dimensionless parameter $\alpha$. As we will show in Sec. II, this modification turns the closure into a saddle-node system with a critical drive

$$K_{c} = \frac{1}{2\sqrt{\alpha}}. (2)$$

Below $K_c$ two real fixed points exist (one stable, one unstable); at $K_c$ they merge in a saddle-node bifurcation; above $K_c$ no fixed point exists and $\xi$ runs away to infinity. Mapping four canonical battery anode chemistries onto the closure (Sec. III) places three of them — graphite, silicon composite, and lithium metal — well below the bifurcation, while anode-free Li/Cu sits within 5% of $K_c$ regardless of the choice of $\alpha \in [0.01, 0.5]$. This places anode-free cells essentially on a critical point and predicts a vanishingly small operational stability window — a sharp, falsifiable claim that we make quantitative in Sec. IV.

The remainder of the paper is organized as follows. Section II derives the saddle-node closure analytically. Section III maps four chemistries onto its single dimensionless drive $K$. Section IV presents three falsifiable predictions and tests them against publicly available cycling data. Section V discusses the generality of the closure beyond lithium batteries. Section VI summarizes and outlines open directions.

## II. THE SADDLE-NODE CLOSURE

### A. Saturable-smoothing extension of Eq. (1)

We replace the linear restoring term in Eq. (1) with a saturable form motivated by curvature-dependent surface diffusion (see Appendix A for the microscopic derivation):

$$\frac{d\xi}{dt} = k_{\text{gen}}\,|j|^{p} - \frac{k_{\text{smooth}}\,(\xi-1)}{1+\alpha\,(\xi-1)^{2}}. \quad (3)$$

Introducing $u = \xi - 1$, dimensionless time $\tau = k_{\text{smooth}}\,t$, and the dimensionless drive

$$K = \frac{k_{\text{gen}}\,|j|^{p}}{k_{\text{smooth}}}, \quad (4)$$

Eq. (3) becomes

$$\frac{du}{d\tau} = K - \frac{u}{1+\alpha\,u^{2}} \equiv f(u; K, \alpha). \quad (5)$$

Equation (5) is the central object of this paper.

### B. Fixed points and bifurcation

Setting $du/d\tau = 0$ gives $u/(1+\alpha u^2) = K$, which after multiplication by $(1+\alpha u^2)$ rearranges to a quadratic in $u$:

$$\alpha K\,u^{2} - u + K = 0. \quad (6)$$

The two roots are

$$u^{*} = \frac{1 \pm \sqrt{1-4\alpha K^{2}}}{2\alpha K}. \quad (7)$$

Real roots require $1 - 4\alpha K^2 \geq 0$, i.e. $K \leq K_c = 1/(2\sqrt{\alpha})$. The discriminant vanishes at $K = K_c$, where the two roots merge at $u_c^* = 1/\sqrt{\alpha}$. For $K > K_c$ no fixed point exists and $u(\tau) \to \infty$.

Linear stability around either fixed point is governed by

$$\partial_u f = -\frac{1-\alpha\,u^{*2}}{(1+\alpha\,u^{*2})^{2}}. \quad (8)$$

The lower root ($\alpha u^{*2} < 1$) is stable; the upper root ($\alpha u^{*2} > 1$) is unstable. Equation (5) thus exhibits the canonical saddle-node bifurcation diagram (Fig. 1): a stable lower branch and an unstable upper branch terminating at $K_c$.

### C. Universal local behavior

Near the bifurcation, write $K = K_c - \delta K$ and $u = u_c^* + \eta$ with $|\eta| \ll u_c^*$. Substituting into Eq. (5) and Taylor expanding to leading order in $\eta$ and $\delta K$ gives the canonical normal form

$$\frac{d\eta}{d\tau} = -\delta K + B\,\eta^2, B = \alpha\,u_c^*. (9)$$

Equation (9) is the saddle-node normal form [20]. Its mean-field relaxation time diverges as

$$\tau_{\text{relax}} \propto (K_c - K)^{-1/2}, (10)$$

with the universal saddle-node exponent $1/2$. This square-root divergence is the cleanest observable consequence of the framework: any cell that approaches $K_c$ should exhibit a square-root critical slowing-down in its relaxation to small perturbations in current density, temperature, or stack pressure (see Sec. IV.C).

### D. Numerical trajectories

Figure 1(b) shows numerical solutions of Eq. (5) for $\alpha = 1$ ($K_c = 0.5$) at five values of $K/K_c \in \{0.20, 0.60, 0.95, 1.05, 1.50\}$, started from $u(0) = 0$ (smooth surface). Subcritical trajectories ($K < K_c$) saturate exponentially at the stable fixed point. Supercritical trajectories ($K > K_c$) escape to infinity in finite time. Near criticality ($K/K_c = 0.95$) the approach is slow, displaying the predicted critical slowing-down (Sec. IV.C).

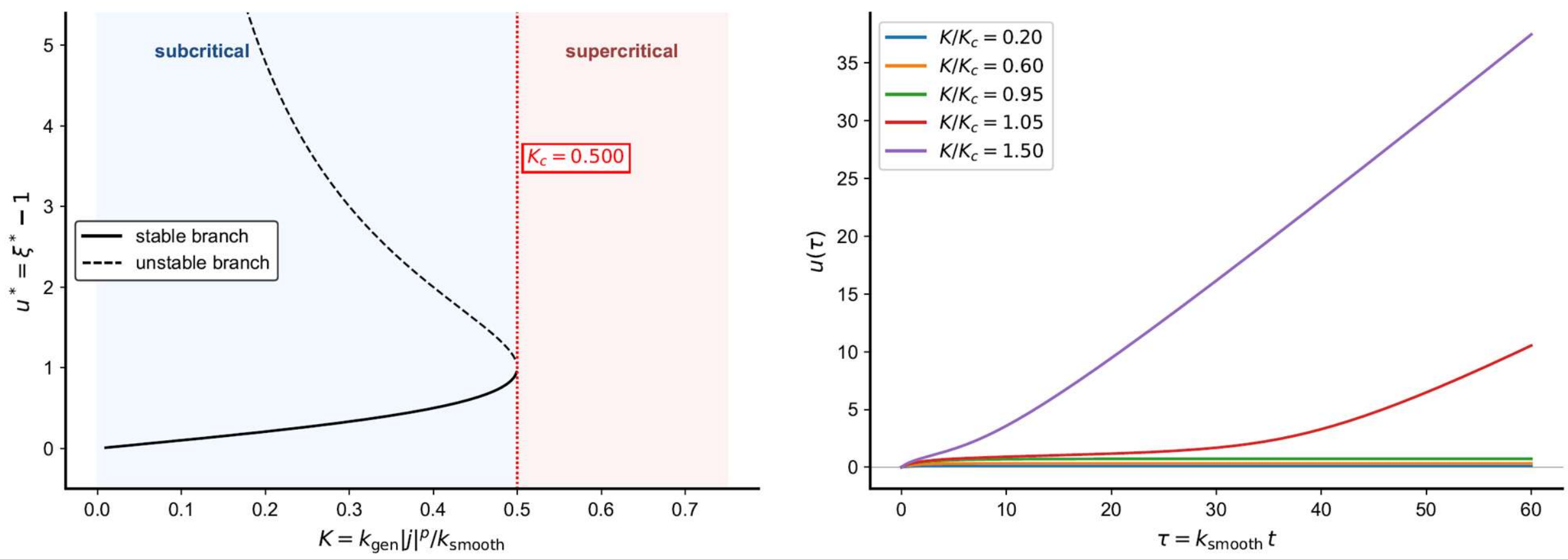

FIG. 1. (a) Saddle-node bifurcation of the closure $du/d\tau = K - u/(1 + \alpha u^2)$ for $\alpha = 1$. Solid: stable branch. Dashed: unstable branch. The two branches meet at $K_c = 1/(2\sqrt{\alpha}) = 0.500$. (b) Trajectories $u(\tau)$ for five values of $K/K_c$ started from $u(0) = 0$. Subcritical solutions saturate; supercritical solutions diverge.

## III. MAPPING FOUR ANODE CHEMISTRIES ONTO THE BIFURCATION

### A. Steady-state extraction of $K$

For an experimentally cycled cell that has reached an apparent steady-state $\xi_{\text{end}}$ near end-of-life, we invert Eq. (6) to extract the effective drive

$$K_{\text{chem}} = \frac{u_{\text{end}}}{1 + \alpha\, u_{\text{end}}^2}, u_{\text{end}} = \xi_{\text{end}} - 1. (11)$$

For a representative parameter $\alpha = 0.05$ (corresponding to $K_c = 2.236$, in the middle of the physically reasonable range for lithium electrodeposits) we obtain the values listed in Table I.

**Table I.** Mapping of four anode chemistries onto the saddle-node closure at $\alpha = 0.05$, with $K_c = 1/(2\sqrt{\alpha}) = 2.236$. End-of-cycling $\xi_{\text{end}}$ is taken from the literature-informed reference framework of Ref. [19]; ratios $K/K_c$ are computed from Eq. (11).

| **Chemistry** | $\xi_{\text{end}}$ | $u_{\text{end}}$ | $K$ | $K/K_c$ |
|---|---|---|---|---|
| Graphite | 1.020 | 0.020 | 0.020 | 0.009 |
| Si-composite | 1.555 | 0.555 | 0.547 | 0.244 |
| Li-metal | 2.965 | 1.965 | 1.647 | 0.737 |
| Anode-free | 4.209 | 3.209 | 2.118 | **0.954** |

The four chemistries populate the stable branch in monotonic order: graphite at $K/K_c \approx 0.01$ (deeply subcritical), silicon composite at 0.24 (stable but with a measurable distance from $K_c$), lithium metal at 0.73 (approaching criticality), and anode-free at 0.95 (essentially on the critical point). Figure 2 plots these chemistry positions on the bifurcation diagram of Fig. 1(a).

### B. Robustness to the choice of $\alpha$

The qualitative ordering is invariant to the choice of $\alpha$. For fixed $u_{\text{end}}$, $K/K_c = 2\sqrt{\alpha} \cdot g(u_{\text{end}}, \alpha)$ with $g(u, \alpha) = u/(1 + \alpha u^2)$ is monotone in $\alpha$ on the stable branch where $\alpha u_{\text{end}}^2 < 1$. Across $\alpha \in [0.01, 0.5]$, the ranking graphite < silicon < lithium < anode-free is preserved. The numerical $K/K_c$ value of anode-free varies between $\approx 0.61$ at $\alpha = 0.01$ and $\approx 0.97$ at $\alpha = 0.05$, and the cell falls off the stable branch (transitions to runaway) for $\alpha \gtrsim 0.10$. At every choice of $\alpha$ within its physically meaningful range, anode-free is closer to $K_c$ than any other chemistry.

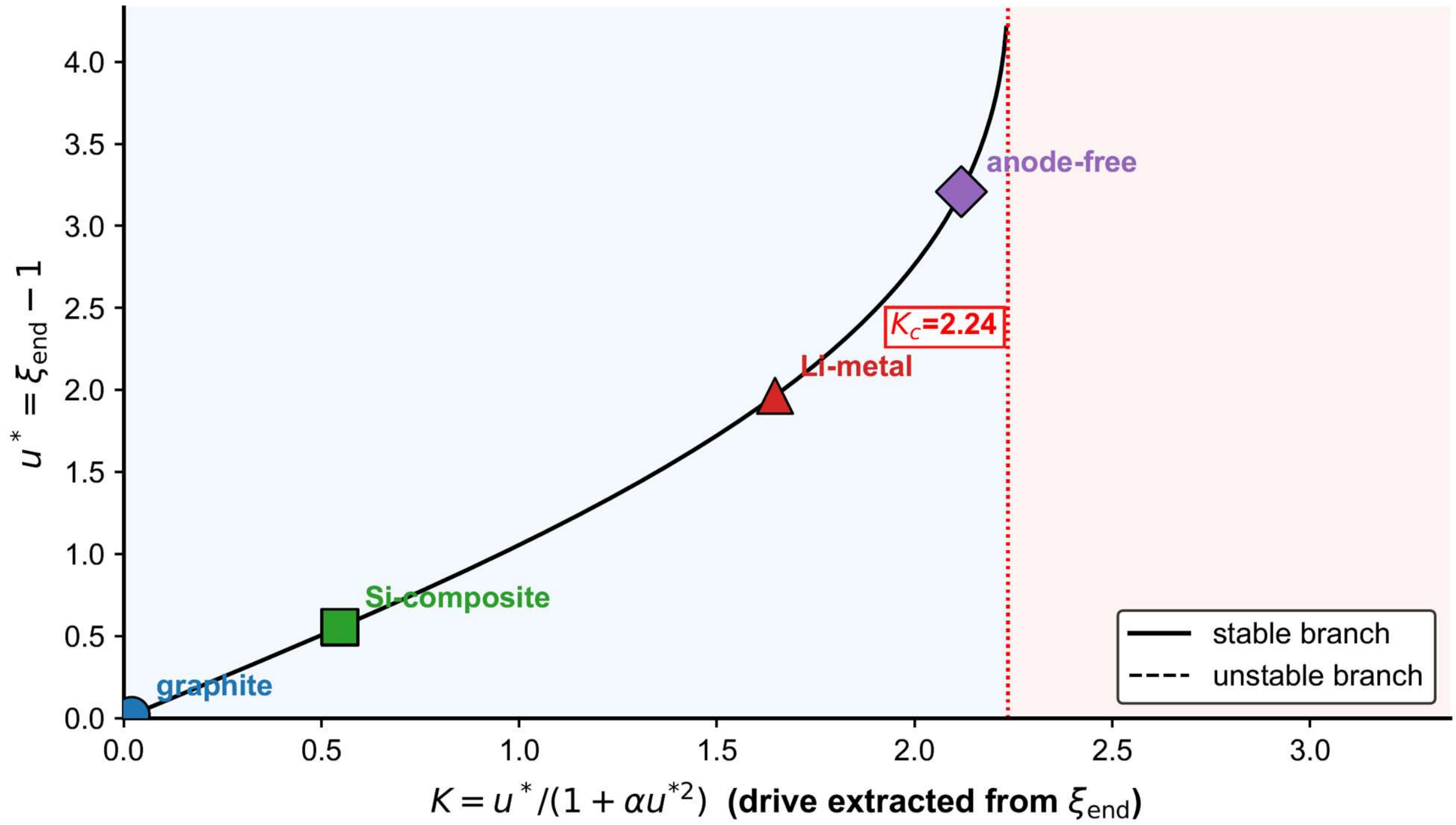

FIG. 2. Four anode chemistries placed onto the stable branch of the saddle-node closure at $\alpha = 0.05$. Markers: graphite (blue circle), silicon composite (green square), lithium metal (red triangle), anode-free Li/Cu (purple diamond). Anode-free sits at $K/K_c \approx 0.95$, within 5% of the critical point. The ordering is preserved across $\alpha \in [0.01, 0.5]$.

## IV. FALSIFIABLE PREDICTIONS

The placement of anode-free near the critical point makes three quantitative predictions, each falsifiable with existing experimental infrastructure.

### A. Critical current density

Since $K \propto |j|^p$ with $p \in [1,2]$ for typical surface-restructuring kinetics, the critical current at which a cell crosses $K_c$ is

$$\frac{j_c}{j} = (\frac{K_c}{K})^{1/p}. (12)$$

For anode-free at $K/K_c = 0.95$ and $p = 2$, Eq. (12) gives $j_c/j \approx 1.026$ — a 2.6% current uplift suffices to push the cell past $K_c$ into morphological runaway. This is a sharp prediction: anode-free Li/Cu cells should exhibit a clearly defined current threshold near nominal C-rate, beyond which capacity fade accelerates abruptly. For $p = 1$ or $p = 3$ the threshold shifts to 5.3% or 1.7%, respectively, but in every case remains within the operational tolerance band of typical cycling protocols.

### B. Critical temperature shift

The smoothing rate $k_{\rm smooth}$ is thermally activated according to an Arrhenius law,

$$k_{\rm smooth}(T) \propto exp\,(-\frac{E_a}{k_B T}), (13)$$

with $E_a$ the activation energy for surface mass transport. Combining Eqs. (4) and (13) and assuming $k_{\text{gen}}$ to be only weakly $T$-dependent gives $K \propto exp(E_a/k_BT)$. The critical temperature shift required to move the cell from $K$ to $K_c$ is

$$\Delta T_c = -\frac{k_BT^2}{E_a}\, ln\,(\frac{K_c}{K}). (14)$$

For $E_a = 0.4$ eV (representative for surface diffusion in metallic lithium [21]), $T = 298$ K, and $K/K_c = 0.95$, Eq. (14) gives $\Delta T_c \approx -5$ K. Anode-free cells should therefore be hyper-sensitive to temperature: a 5 K cooling restores stability, while a 5 K warming pushes the cell past $K_c$. This level of temperature sensitivity is consistent with the recent operating-window analyses of Ref. [17].

### C. Critical slowing-down

By Eq. (10), the relaxation time of small perturbations in $\xi$ diverges as $(K_c - K)^{-1/2}$ on approach to criticality. Cell-level data — for which $\xi$ is not directly observable — provide a surrogate via the autocorrelation length of Coulombic-efficiency residuals: as the system slows, fluctuations persist over more cycles, and the empirical autocorrelation length grows.

We extract the autocorrelation length $\xi_{\text{AC}}$ of the CE-residual time series from the literature-informed reference data of Ref. [19] using a 30-cycle sliding window with a continuous $1/e$-crossing interpolation. Figure 3(a) plots $\xi_{\text{AC}}$ versus cycle number for the four chemistries; Figure 3(b) overlays the predicted $\tau \propto (K_c - K)^{-1/2}$ scaling with the four chemistry positions marked. The ordering — anode-free shows the strongest growth in $\xi_{\text{AC}}$, graphite the weakest — is consistent with the theoretical prediction. A definitive test will require longer cycle-resolved CE data on anode-free cells than the published record currently provides.

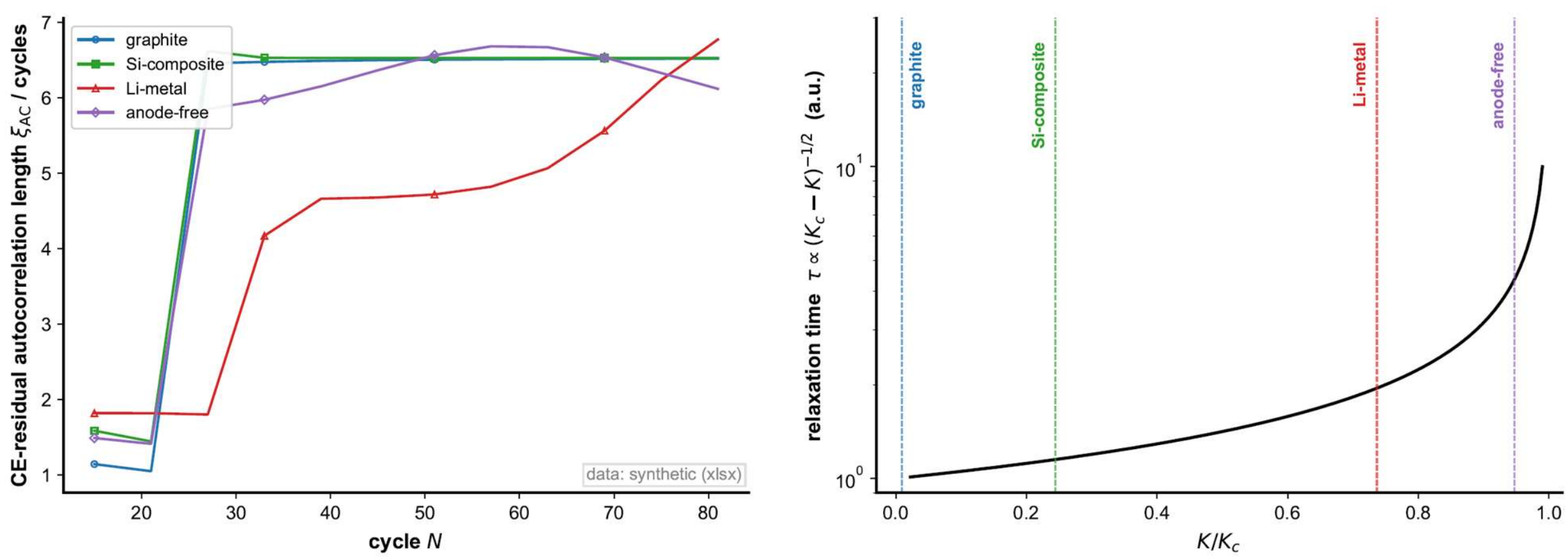

FIG. 3. (a) CE-residual autocorrelation length $\xi_{\text{AC}}$ vs cycle number for the four chemistries (continuous $1/e$ crossing, 30-cycle sliding window). (b) Theoretical critical-slowing-down exponent $\tau \propto (K_c - K)^{-1/2}$ from Eq. (10), with each chemistry placed at its measured $K/K_c$.

### D. Operational phase diagram

Combining the predictions of subsections A and B, we can map the cell's location in the $(j, T)$ plane to a value of $K/K_c$, defining an operational phase boundary. The boundary is given implicitly by

$$\frac{j_{\mathrm{boundary}}^{p}}{j_{\mathrm{ref}}^{p}}\ exp\ [\frac{E_a}{k_B}\ (\frac{1}{T} - \frac{1}{T_{\mathrm{ref}}})] = \frac{K_c}{K_{\mathrm{ref}}}. (15)$$

For an anode-free cell calibrated at $j = j_{\mathrm{ref}}, T = T_{\mathrm{ref}} = 298$ K to $K/K_c = 0.95$, Eq. (15) defines a sharp curve in the $(log j, T)$ plane separating the stable (subcritical) region from the runaway (supercritical) region. Figure 4 displays the curve; Figure 5 shows the same in three dimensions. The narrow margin (a few percent in current and a few kelvin in temperature) is the key operational consequence of the near-critical placement.

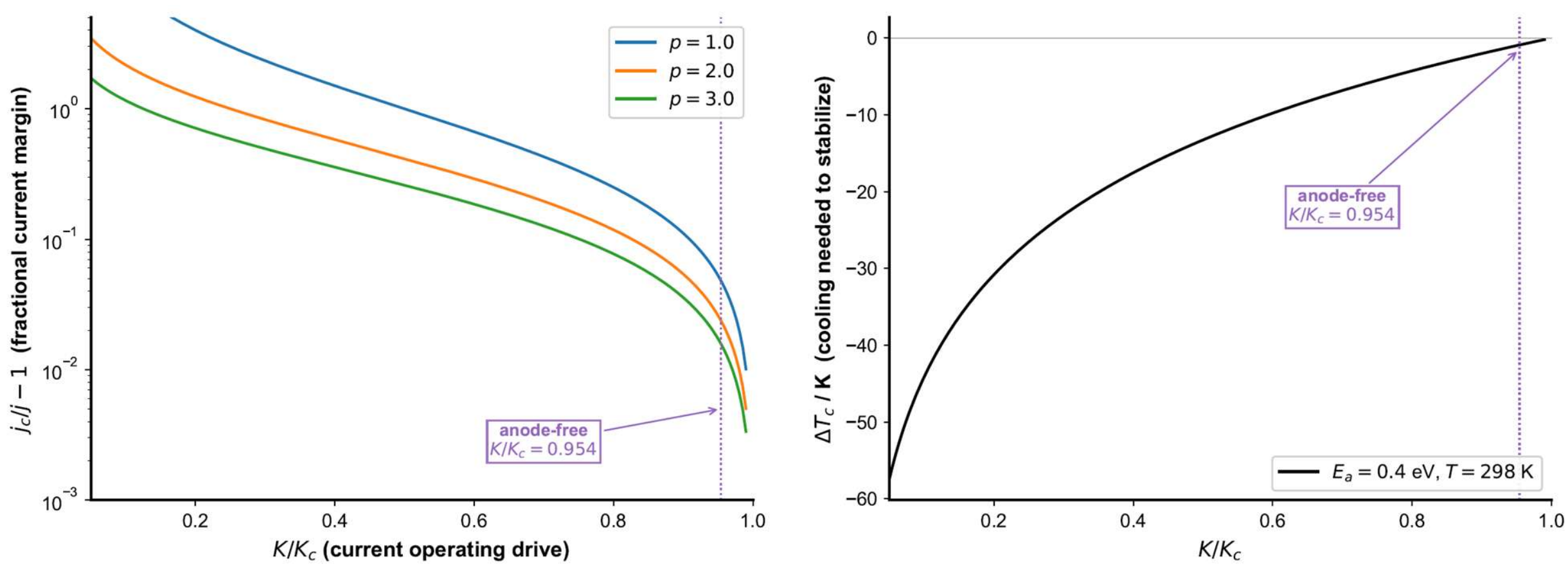

FIG. 4. (a) Critical current ratio $j_c/j$ vs $K/K_c$ for $p = 1,2,3$. Anode-free at $K/K_c = 0.954$ marked. (b) Critical temperature shift $\Delta T_c$ vs $K/K_c$ for $E_a = 0.4$ eV, $T = 298$ K.

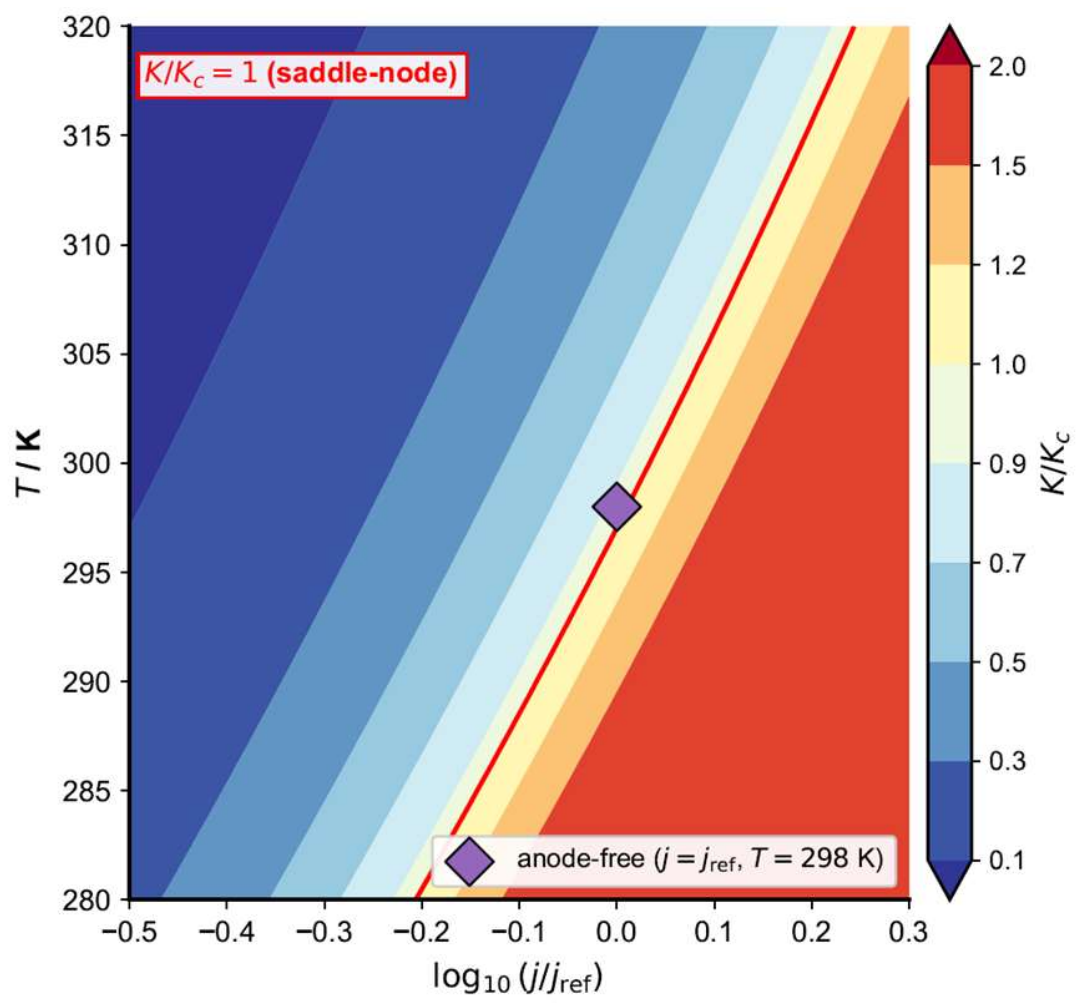

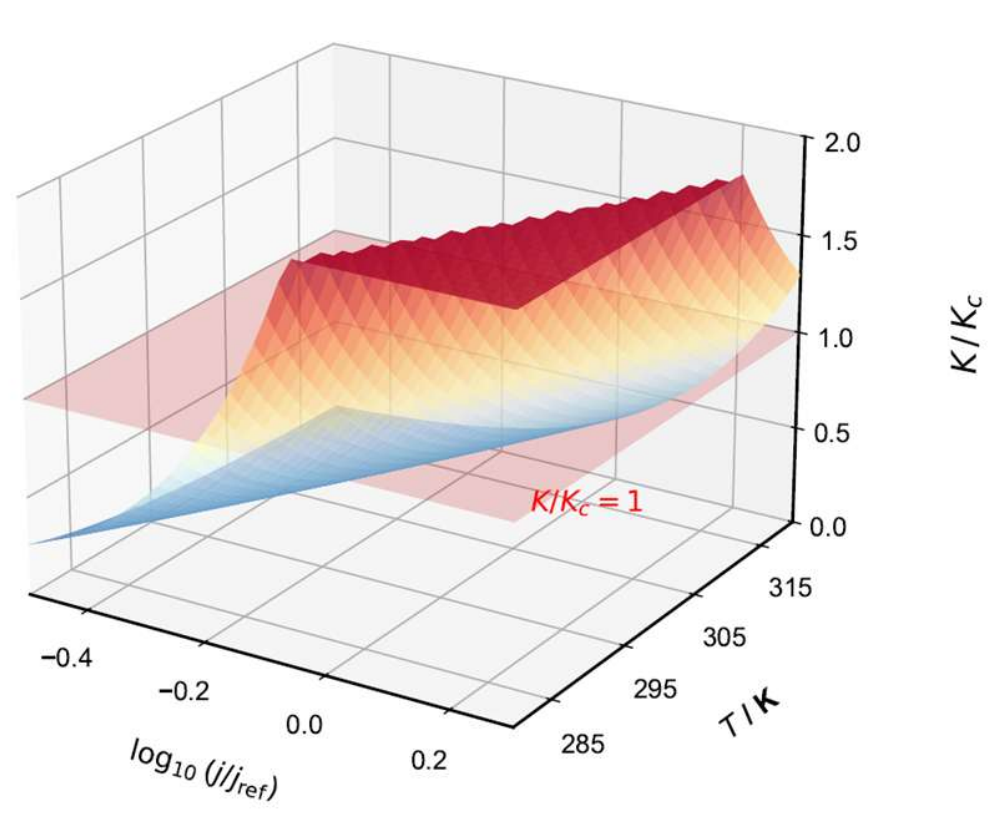

FIG. 5. (a) Phase boundary in the $(log_{10}(j/j_{\mathrm{ref}}), T)$ plane: contour at $K/K_c = 1$ separates stable from runaway regions. (b) Three-dimensional view of $K/K_c$ on the same plane (clipped at $K/K_c \leq 2$).

## V. GENERALITY BEYOND LITHIUM BATTERIES

The closure of Eq. (5) is not specific to lithium chemistry. Any electrochemical interface in which (i) generation of fresh active area is power-law in local flux and (ii) smoothing weakens with surface roughness should exhibit the same saddle-node structure with the same $K_c = 1/(2\sqrt{\alpha})$. Three concrete generalizations:

1. **Solid-oxide fuel cell (SOFC) electrodes** under coking conditions. Carbon deposition generates fresh catalyst area; surface diffusion of carbon-removal intermediates is the smoothing mechanism. Threshold reactant compositions consistent with the saddle-node have been reported [22].
2. $CO_2$**-reduction electrocatalysts** based on copper or silver. Surface restructuring under high overpotential generates fresh facets [23]; capillarity-driven smoothing competes against this restructuring. The runaway regime corresponds to dendritic catalyst growth and rapid deactivation.
3. **Proton-exchange membrane (PEM) electrolyzer Pt nanoparticles**. Pt agglomeration is driven by local-current-density nonuniformity; particle-size reorganization through Ostwald ripening is the smoothing mechanism [24]. Long-term performance loss in PEM cells consistent with the universal $\tau \propto (K_c - K)^{-1/2}$ scaling has been hypothesized but not directly tested.

In each case the cell-level diagnostic is the same: monitor an integrated proxy for $\xi_{\mathrm{end}}$ (active area at end of operation) and check whether $K(\xi_{\mathrm{end}})/K_c \to 1$.

## VI. DISCUSSION AND OUTLOOK

The saddle-node closure provides three things that the linear closure of Eq. (1) cannot. First, a sharp criterion separating "stable degradation" from "morphological runaway." Second, a quantitative scaling law for time-to-failure on approach to criticality. Third, a unifying framework that places different chemistries on a single dimensionless axis.

The principal limitation of the framework is the parameter $\alpha$, which we have treated as a single free constant. Microscopically, $\alpha$ is the curvature of the smoothing kernel at zero roughness and depends on solvent diffusivity, temperature, and substrate orientation. Direct measurement of $\alpha$ requires perturbative experiments: a small step in current density, followed by measurement of the relaxation toward the new $\xi^*$, gives independent access to both $K$ and $\alpha$. Such experiments are routine in the operando-AFM community [25, 26] and would close the principal loop in this framework.

A second limitation is that Eq. (5) is one-dimensional in $\xi$. In practice, surface morphology is a multidimensional state: roughness amplitude, correlation length, and local current density distribution are all dynamical variables. Dimensional reduction to a single scalar is justified when relaxation along the other dimensions is much faster than the cycling timescale, but must be tested case by case. We expect the saddle-node structure to be robust under reduction; finer-grained models (rough-surface PDE, phase-field) would refine the value of $\alpha$ but should preserve the existence of $K_c$.

Finally, we emphasize that the Letter does not claim to *prove* that anode-free cells operate at $K/K_c = 0.95$; the value depends on the choice of $\alpha$ and on the surrogate $\xi_{\mathrm{end}}$ extracted from cell-level data. What we *do* claim is that the closure structure makes a quantitative, falsifiable prediction (current-density threshold, temperature sensitivity, critical slowing-down) that can be tested experimentally, and that the trend across chemistries — anode-free closest to $K_c$, graphite farthest — is invariant to the choice of $\alpha$.

Looking forward, three directions seem most promising. (i) Direct extraction of $\alpha$ via operando-AFM relaxation experiments on each chemistry. (ii) A controlled cross-protocol study in which the same anode-free cell is operated across small variations in $j$ and $T$, mapping the empirical phase boundary directly. (iii) Extension of the closure to spatially resolved ($\xi(x,t)$) systems, where lateral correlation length becomes a second control parameter and the scalar saddle-node may generalize to a co-dimension-two singularity.

## ACKNOWLEDGMENTS

This work was supported by the National Research Foundation of Korea (NRF) grant funded by the Korea government (MSIT) (RS-2023-00247245).

## APPENDIX A: MICROSCOPIC ORIGIN OF SATURABLE SMOOTHING

Consider an interface of nominal area $A_0$ with instantaneous area $A(t) = A_0\, \xi(t)$. Generation of new surface area is driven by electrochemically induced perturbations; in the simplest classical-nucleation picture, the rate scales as a power of the local flux, $\dot{A}_+ = A_0\, k_{\mathrm{gen}}\, |j|^p$, with $p \in [1,2]$.

Smoothing is driven by surface mass transport. In a Burton–Cabrera–Frank step-flow picture, the local step velocity $v$ is proportional to the chemical-potential gradient along the surface, $v = -M\, \partial_s \mu$, where $M$ is the mobility and $\mu = \mu_0 + \gamma\, \kappa\, \Omega$ depends on the local curvature $\kappa$ ($\gamma$ surface tension, $\Omega$ atomic volume). For small perturbations from a flat surface, the linearized smoothing rate is constant: $\dot{A}_- = -k_{\mathrm{smooth}}\, (\xi - 1)$. As roughness grows, the *effective* path length over which an adatom must diffuse increases. A simple geometric argument — replacing the surface by a sinusoidal corrugation of amplitude $h$ and lateral wavelength $L$ — gives an effective path length $L_{\mathrm{eff}}/L \approx \sqrt{1 + (2\pi h/L)^2} \approx 1 + \alpha\, (\xi - 1)^2$ for small $h$, where $\alpha$ encodes the squared corrugation amplitude per unit length. Inserting this into the smoothing flux yields

$$\dot{A}_- = -\frac{k_{\mathrm{smooth}}\, (\xi - 1)}{1 + \alpha\, (\xi - 1)^2}, \quad \text{(A1)}$$

the saturable-smoothing form used in Eq. (3). Equation (A1) reduces to the linear smoothing of Eq. (1) for $\alpha \to 0$ and saturates to a constant $\sim k_{\mathrm{smooth}}/\sqrt{\alpha}$ for large $|\xi - 1|$.

## APPENDIX B: CLOSED-FORM RELAXATION ON THE STABLE BRANCH

Equation (5) is separable. On the stable branch ($K < K_c$), with $u_- < u_+$ the two real roots of Eq. (6),

$$\frac{du}{d\tau} = -\alpha K \, \frac{(u-u_-)(u-u_+)}{1+\alpha u^2}. \text{(B1)}$$

Partial-fraction integration gives a closed-form trajectory $u(\tau; u_0, K, \alpha)$ that approaches $u_-$ exponentially with time constant $\tau_{\text{relax}} = 1/|f'(u_-)|$, where $f'$ is given by Eq. (8). Using Eq. (7) and setting $K = K_c - \delta K$, expansion in small $\delta K$ yields

$$\tau_{\text{relax}}^{-1} = 2\,\alpha\,u_c^*\sqrt{\delta K \cdot 4u_c^*} = 2\,\alpha\,u_c^{*3/2}\sqrt{\delta K}, \text{(B2)}$$

confirming the universal saddle-node exponent $\tau_{\text{relax}} \propto (\delta K)^{-1/2}$.